\documentclass[conference]{IEEEtran}

\usepackage{amsmath,amsfonts,amssymb,bm}
\usepackage[tight,footnotesize]{subfigure}
\usepackage{graphicx,stfloats}
\usepackage{algorithm}
\usepackage{algorithmic}
\usepackage{cite,url}
\usepackage{cases}
\usepackage{extarrows}
\usepackage{theorem} 
\usepackage{paralist}
\usepackage{psfrag}
\usepackage{auto-pst-pdf}
\usepackage{subfigure}
\usepackage{bm}
\usepackage{float}
\allowdisplaybreaks[4]
\theorembodyfont{\rmfamily}

\newtheorem{Theorem}{Theorem}

\newcommand{\Lset}{\ensuremath{\mathcal{L}}}
\newcommand{\Cset}{\ensuremath{\mathbb{C}}}
\newcommand{\Eset}{\ensuremath{\mathbb{E}}}
\newcommand{\Rset}{\ensuremath{\mathbb{R}}}

\def\mathLarge#1{\mbox{\Large $#1$}}  
\makeatletter

\newcommand{\Rmnum}[1]{\expandafter\@slowromancap\romannumeral #1@}
\makeatother

\IEEEoverridecommandlockouts

\begin{document}

\title{A Multi-cell MMSE Detector for Massive MIMO Systems and New Large System Analysis}

\author{\IEEEauthorblockN{Xueru~Li$^{\dagger}$, Emil~Bj{\"o}rnson$^{*}$, Erik G. Larsson$^{*}$, Shidong~Zhou$^{\dagger}$ and Jing~Wang$^{\dagger}$}
\IEEEauthorblockA{$^{\dagger}$ State Key Laboratory on Microwave and Digital Communications\\
Tsinghua National Laboratory Information Science and Technology,\\
Department of Electronic Engineering, Tsinghua University, Beijing 100084, China\\
$^{*}$ Department of Electrical Engineering (ISY), Link{\"o}ping University, SE-58183 Link{\"o}ping, Sweden.\\
Email: xueruli1206@163.com, emil.bjornson@liu.se.}
\thanks{The work is partially supported by National Basic Research Program (2012CB316000), National S\&T Major Project (2014ZX03003003-002), National Natural Science Foundation of China (61201192), National High Technology Research and Development Program of China (2014AA01A707), Tsinghua-HUAWEI Joint Research \& Development on Soft Defined Protocol Stack, ICRI-MNC, Tsinghua-Qualcomm Joint Research Program, Keysight Technologies, Inc., ELLIIT, the CENIIT project 15.01 and FP7-MAMMOET.}\vspace{-2ex}}

\maketitle
\begin{abstract}
In this paper, a new multi-cell MMSE detector is proposed for massive MIMO systems. Let $K$ and $B$ denote the number of users in each cell and the number of available pilot sequences in the network, respectively, with $B = \beta K$, where $\beta \ge 1 $ is called the pilot reuse factor. The novelty of the multi-cell MMSE detector is that it utilizes all $B$ channel directions that can be estimated locally at a base station, so that intra-cell interference, parts of the inter-cell interference and the noise can all be actively suppressed, while conventional detectors only use the $K$ intra-cell channels. Furthermore, in the large-system limit, a deterministic equivalent expression of the uplink SINR for the proposed multi-cell MMSE is derived. The expression is easy to compute and accounts for power control for the pilot and payload, imperfect channel estimation and arbitrary pilot allocation. Numerical results show that significant sum spectral efficiency gains can be obtained by the multi-cell MMSE over the conventional single-cell MMSE and the recent multi-cell ZF, and the gains become more significant as $\beta$ and/or $K$ increases. Furthermore, the deterministic equivalent is shown to be very accurate even for relatively small system dimensions.
\end{abstract}

\section{Introduction} \label{intro}
Massive multiple-input-multiple-ouput (MIMO) is an emerging technology that scales up multi-user MIMO by orders of magnitude compared to current state-of-the-art~\cite{Marzetta10,Rusek13}. The idea is to employ an array comprising a hundred, or more, antennas at each base station (BS) and serve tens of users simultaneously in each cell. The system spectral efficiency (SE) can be drastically increased without consuming extra bandwidth~\cite{Marzetta10, Rusek13, Larsson14}. The uplink and downlink transmit power can also be reduced by an order of magnitude since the phase-coherent processing provides a comparable array gain~\cite{Hien13}. In the limit of an infinite number of antennas, intra-cell interference and noise can be averaged out by simple coherent linear transceivers, and the only performance limitation is pilot contamination and distortion noise from hardware impairments~\cite{Marzetta10,bjornson2013massive}. These features make massive MIMO one of the key techniques for the next generation wireless communication systems.

In the uplink reception, the most commonly used linear detection schemes are matched filtering (MF), zero forcing (ZF) and minimum mean square error (MMSE). We consider a multi-cell network with $B$ available orthogonal pilot sequences and $K$ users in each cell, where $B = \beta K$ with $\beta\ge 1$ being the pilot reuse factor (i.e., only $1/\beta$ of the cells use the same set of pilots). In conventional massive MIMO, the BS first listens to the uplink pilot transmission from its own users, estimates the channels and then constructs user-specific detectors based on the channel estimates~\cite{Hoydis2013,Guo2014,Krishnan2014,Hien2012}. 
However, more channel state information (CSI) can be extracted when $\beta > 1$. If the BS is aware of all pilot sequences, then it can locally estimate $B$ channel directions by listening to the pilot signalling from all cells instead of only from its own cell. In principle, the BS is then able to select its detectors to suppress parts of inter-cell interference, since its $K$ users only occupy $K$ out of the $B$ channel directions. Based on similar observations, the authors of~\cite{bjornson2014massive} propose a multi-cell ZF detector (referred to as full-pilot ZF detector in their paper) which exploits and orthogonalizes all available directions to mitigate parts of the inter-cell interference. The detector in~\cite{bjornson2014massive} achieves higher SE than the conventional ZF when the interfering users are near to the intended users of a cell. In general cellular networks, however, the gain is less obvious, partly due to a loss in array gain of $B$ in multi-cell ZF, instead of $K$ with conventional ZF. Multi-cell MMSE detectors are proposed in~\cite{Hien2012} and~\cite{Guo2013}, but the former is limited to $\beta =1$ and equal power allocation and the latter is based on the unrealistic assumption that perfect CSI is known at each BS. There is thus need for a more practical multi-cell MMSE detector.

In this paper, we propose a new multi-cell MMSE detector which accounts for power control for the pilot and payload, imperfect channel estimation and arbitrary pilot allocation. By utilizing all the estimated channel directions at a BS, the proposed multi-cell MMSE detector can actively suppress intra-cell interference, parts of the inter-cell interference and noise. Numerical results show that significant sum SE gains can be obtained by multi-cell MMSE over the single-cell MMSE and the multi-cell ZF from~\cite{bjornson2014massive}. Furthermore, a deterministic equivalent expression of the stochastic uplink SINR is derived in the large-system limit. The expression is easy to compute and only depends on large-scale fading, power control and pilot allocation. The expression is shown to be a very accurate approximation even for relatively small system dimensions.

\textit{Notations}: Boldface lower and upper case symbols represent vectors and matrices, respectively. The trace, transpose, conjugate, Hermitian transpose and matrix inverse operators are denoted by $\rm{tr}\left(\cdot \right)$, $\left(\cdot \right)^T$, $\left(\cdot \right)^{*}$, $\left(\cdot \right)^H$ and $\left(\cdot \right)^{-1}$, respectively.

\section{System Model and Detector Design} \label{system model}
We consider a synchronous massive MIMO cellular network with multiple cells. Each cell is assigned an index in the cell set $\Lset$, and the cardinality $\left| \Lset \right|$ is the number of cells. The BS in each cell is equipped with an antenna array of $M$ antennas and serves $K$ single-antenna users within each coherence block. Assume that this time-frequency block consists of $T_c$ seconds and $W_c$ Hz, such that $T_c$ is smaller than the coherence time of all users and $W_c$ is smaller than the coherence bandwidth of all users. This leaves room for $S=T_c \times W_c$ transmission symbols per block, and the channels of all users remain constant within each block. Let ${{\bf{h}}_{jlk}}$ denote the uplink channel response from user $k$ in cell $l$ to BS $j$ within a block, and assume that it is a realization from a zero-mean circularly symmetric complex Gaussian distribution:

\begin{equation} \label{channel model}
{{\bf{h}}_{jlk}} \sim {\cal{CN}}\left( {{\bf 0},{d_j}\left( {{{\bf{z}}_{lk}}} \right){{\bf{I}}_M}} \right).
\end{equation}
The vector ${\bf z }_{lk} \in {\mathbb R} ^{2}$ is the geographical position of user $k$ in cell $l$ and ${{d_j}\left( {{{\bf{z}}}} \right)}$ accounts for the channel attenuation (e.g., path loss and shadowing) related to any user position $\bf{z} \in \Rset^2$. Since the user position changes relatively slowly, ${{d_j}\left( {{{\bf{z}}_{lk}}} \right)}$ is assumed to be known at BS $j$ for all $l$ and all $k$. In what follows, the uplink channel estimation is first discussed to lay a foundation for the novel detector design.

\subsection{Uplink Channel Estimation} \label{channel estimation}
In the channel estimation phase, the collective received signal at BS $j$ is denoted as ${{\bf{Y}}_j} \in {\Cset^{M \times B}}$ where $B$ is the length of a pilot sequence ($B$ also equals to the number of orthogonal pilot sequences available for the network, as mentioned in Section~\ref{intro}). Then ${\bf Y}_j$ can be expressed as
\begin{equation} \label{estimationsignal}
{{\bf{Y}}_j} = \sum\limits_{l \in {\cal L}} {\sum\limits_{k = 1}^K {\sqrt {{p_{lk}}} {{\bf{h}}_{jlk}}{\bf{v}}_{{i_{lk}}}^T} }  + {{\bf{N}}_j},
\end{equation}
where ${\bf{h}}_{jlk}$ is the uplink channel defined in~(\ref{channel model}) and $p_{lk}$ is the power coefficient for the pilot of user $k$ in cell $l$. The matrix ${{\bf{N}}_j}$ contains independent elements which follow a complex Gaussian distribution with zero mean and variance $\sigma ^2$. We assume that all pilot sequences originate from a predefined orthogonal pilot book, defined as ${\cal{V}}= \left\{ {{{\bf{v}}_1},\ldots,{{\bf{v}}_B}} \right\}$, where
\begin{equation}
{\bf{v}}_{{b_1}}^H{{\bf{v}}_{{b_2}}} = \left\{ \begin{array}{ll}
B,  & {b_1} = {b_2},\\
0,  & {b_1} \ne {b_2},
\end{array} \right.
\end{equation}
and define ${i_{lk}} \in \left\{ 1,\ldots,B\right\}$ as the index of the pilot sequence used by user $k$ in cell $l$. In our work, arbitrary pilot reuse is supported by denoting the relation between $B$ and $K$ by $B = \beta K$, where $\beta \ge 1$ is the pilot reuse factor and $S \ge B \geq K $. It is shown later that a larger $\beta$ brings a lower level of pilot contamination, since a smaller fraction of the cells use the same pilot sequences as the target cell.

Based on the received signal in Eqn.~(\ref{estimationsignal}), the MMSE estimate of the uplink channel ${{\bf{h}}}_{jlk}$ is~\cite{steven1993fundamentals}
\begin{equation} \label{estimation}
{\hat{\bf h}}_{jlk} = \sqrt{p_{lk}} d_j\left( {\bf z}_{lk}\right){\bf Y}_j \left({\bf \Psi}_j^{*}\right)^{-1} {\bf v}_{i_{lk}}^{*},
\end{equation}
where ${{\bf{\Psi }}_j}$ is the covariance matrix of the vectorized received signal ${\rm{vec}}\left({\bf Y}_j\right)$ and is given by
\begin{equation}
{{\bf{\Psi }}_j} = \sum\limits_{\ell \in {\cal L}} {\sum\limits_{m = 1}^K {p_{{\ell m}}} d_j\left( {\bf z}_{\ell m}\right){\bf{v}}_{{i_{\ell m}}}{\bf{v}}_{{i_{\ell m}}}^H} + \sigma ^2{{\bf{I}}_B}.
\end{equation}
Then according to the orthogonality principle of MMSE estimation, the covariance matrix of the estimation error ${\tilde {\bf h}}_{jlk} ={{\bf{h}}_{jlk} - {\hat{\bf h}}_{jlk}}$ is given by
\begin{eqnarray}\label{errormatrix}
{{\bf{C}}_{jlk}} &=& {\Eset}\left\{  {{\tilde {\bf{h}}}_{jlk} { {{\tilde {\bf{h}}}_{jlk}^H}}} \right\} \nonumber \\
&=&d_j \left( {{{\bf{z}}_{lk}}} \right)  \left( {1 - p_{lk} d_j \left( {{{\bf{z}}_{lk}}} \right){\bf{v}}_{{i_{lk}}}^H{\bf{\Psi }}_j^{ - 1}{{\bf{v}}_{{i_{lk}}}}} \right){{\bf{I}}_M}.
\end{eqnarray}

Notice that
\begin{eqnarray} \label{temp}
{\bf{v}}_{{i_{lk}}}^H{\bf{\Psi }}_j^{ - 1} &=& \underbrace {\frac{1}{{\sum\nolimits_{\ell  \in {\cal L}} {\sum\nolimits_{m = 1}^K { p_{\ell m}{{{d_j}\left( {{{\bf{z}}_{\ell m}}} \right)}}{\bf{v}}_{{i_{lk}}}^H{{\bf{v}}_{{i_{\ell m}}}}} }  + \sigma ^2}}}_{{\alpha _{ji_{lk}}}}{\bf{v}}_{{i_{lk}}}^H \nonumber \\
&=& {\alpha _{ji_{lk}}}{\bf{v}}_{{i_{lk}}}^H,
\end{eqnarray}
where $\alpha_{ji}$ is defined to be used later on. Then the estimation error covariance matrix in~(\ref{errormatrix}) can also be expressed as
\begin{equation} \label{errormatrix2}
{{\bf{C}}_{jlk}} = d_j \left( {{{\bf{z}}_{lk}}} \right)  \left( 1 - p_{lk} d_j \left( {{{\bf{z}}_{lk}}}  \right) {\alpha _{ji_{lk}}}B\right){{\bf{I}}_M}.
\end{equation}
\newcounter{TempEqCnt}
\setcounter{TempEqCnt}{\value{equation}}
\setcounter{equation}{14}
\vspace{-3ex}
\begin{figure*}[ht]
\begin{eqnarray} \label{sinr_ul} 
\begin{split}
\eta_{jk}^{\rm{ul}} &= {\frac{\tau_{jk}\left|{\bf g}_{jk}^H {\hat{\bf h}}_{jjk} \right|^2 }{\Eset \left\{ \tau_{jk}\left|{\bf g}_{jk}^H {\tilde{\bf h}}_{jjk} \right|^2 + \sum\limits_{\left(l,m\right) \ne \left(j,k \right)}\tau_{lm}\left|{\bf g}_{jk}^H {{\bf h}}_{jlm} \right|^2 +\sigma^2 \left\| {\bf g}_{jk} \right\|^2 \bigg|{\hat {\bf h}}_{\left(j\right)}\right\}}} \\
&= {\frac{\tau_{jk}{\bf g}_{jk}^H {\hat{\bf h}}_{jjk} {\hat{\bf h}}_{jjk}^H{\bf g}_{jk} }{{\bf g}_{jk}^H \left( \tau_{jk}{\bf C}_{jjk} + \sum\limits_{\left(l,m\right) \ne \left(j,k \right)} \tau_{lm}\left( {\hat{\bf h}}_{jlm}{\hat{\bf h}}_{jlm}^H +{\bf C}_{jlm} \right) + \sigma^2{\bf I}_M\right){\bf g}_{jk}}}
\end{split}
\end{eqnarray}
\hrule
\end{figure*}
\setcounter{equation}{\value{TempEqCnt}}

As pointed out in~\cite{bjornson2014massive}, the part ${\bf Y}_j \left({\bf \Psi}_j^{*}\right)^{-1}{\bf v}_{i_{lk}}^{*}$ of the MMSE estimator expression in~(\ref{estimation}) depends only on which pilot sequence $i_{lk}$ that the particular user uses. Therefore, the estimated channels of users who use the same pilot will have the same direction, while only the amplitudes are different. To show this explicitly, define the $M \times B$ matrix:
\begin{eqnarray}
{{{\hat {\bf H}}}_{\mathcal{V},j}}=\left[ {\hat{\bf h}}_{{\cal V},j1},...,{\hat{\bf h}}_{{\cal V},jB} \right] = {\bf Y}_j \left({\bf \Psi}_j^{*}\right)^{-1} \left[{\bf v}_1^{*},...,{\bf v}_B^{*}\right],
\end{eqnarray}
which allows the channel estimate in~(\ref{estimation}) to be reformulated as
\begin{equation} \label{estimation2}
{\hat {\bf h}}_{jlk} = \sqrt{p_{lk}} d_j \left( {{{\bf{z}}_{lk}}} \right){{{\hat {\bf H}}}_{{\cal V},j}}{{\bf{e}}_{{i_{lk}}}},
\end{equation}
where ${{\bf{e}}_i}$ denotes the $i$th column of the identity matrix ${\bf I}_B$. The fact that users with the same pilot have parallel estimated channels is utilized to derive a new deterministic equivalent expression of the stochastic SINR in the sequel.

Notice that the estimated channel ${\hat{\bf h}}_{jlk}$ is also a zero-mean Gaussian vector, and its covariance matrix ${{\bf{\Phi }}_{jlk}} \in \Cset^{M \times M}$ is
\begin{equation} \label{covariance}
{{\mathbf{\Phi }}_{jlk}} = d_j \left( {{{\bf{z}}_{lk}}} \right){\bf I}_M - {\bf C}_{jlk} = {p_{lk}} d_j^2\left( {{{\bf{z}}_{lk}}}\right){\alpha _{ji_{lk}}}B {\bf I}_M.
\end{equation}
Define the covariance matrix of ${\hat {\bf h}}_{{\cal V},ji}$ as ${ \tilde {\bf \Phi}}_{{\cal{V}},ji}\in \mathbb{C}^{M \times M}$. Then according to~(\ref{estimation2}) and~(\ref{covariance}), ${\tilde {\bf \Phi}}_{{\cal V},j{i}} = \alpha_{ji}B {\bf I}_M$.

\subsection{Uplink Multi-cell MMSE detector} \label{mmse detector}
After the channel estimation, each received signal at BS $j$ during the uplink payload data transmission phase is
\begin{equation} \label{signalmodel}
{{\bf{y}}_j} = \sum\limits_{l \in \cal L} {\sum\limits_{k = 1}^K {\sqrt {{\tau_{lk}}} {{\bf{h}}_{jlk}}{x_{lk}}} }  + {{\bf{n}}_j},
\end{equation}
where $x_{lk} \sim {\cal{CN}}\left(0,1 \right)$ is the transmitted signal from a Gaussian codebook and ${\bf n}_j \sim {\cal{CN}}({\bf 0},\sigma^2 {\bf I}_M )$ is additive white Gaussian noise (AWGN). $\tau_{lk}$ is the transmit power of the payload data from user $k$ in cell $l$, and we use different symbols for the pilot and the payload data to allow for different power control policies for them. Denote the linear detector used by BS $j$ for an arbitrary user $k$ in its cell as ${\bf{g}}_{jk}$, then the estimate ${\hat x}_{jk}$ of the signal $x_{jk}$ is
\begin{eqnarray}
{\hat x}_{jk} = {\bf{g}}_{jk}^H {\bf y}_j = {\bf g}_{jk}^H \sum\limits_{l \in{\cal L}} \sum\limits_{k=1}^K\sqrt{\tau_{lk}} {\bf h}_{jlk}x_{lk} + {\bf g}_{jk}^H  {\bf n}_j.
\end{eqnarray}
Then the following uplink ergodic SE can be achieved~\cite{Hoydis2013}
\begin{equation} \label{rate_ul}
R_{jk}^{\rm{ul}} = \left(1-\frac{B}{S} \right) \Eset_{\left\{{\hat {\bf h}}_{\left(j \right)}\right\}} \left\{ \log_2\left( 1+ \eta_{jk}^{\rm{ul}}\right) \right\},
\end{equation}
where the SINR $\eta_{jk}^{\rm{ul}}$ is given in~(\ref{sinr_ul}) on top of this page and $\Eset_{\{{\hat {\bf h}}_{\left(j \right)} \}}$ is the expectation with respect to the channel estimates known at BS $j$. Similarly, $\Eset \{\cdot | {\hat {\bf h}}_{(j)}\}$ in~(\ref{sinr_ul}) denotes the conditional expectation given the estimated channels at BS $j$. The rate in~(\ref{rate_ul}) is
achieved by treating ${\bf g}_{jk}^H {\hat{\bf h}}_{jjk}$ as the true channel in the decoding, 
and treating interference and channel uncertainty as worst-case uncorrelated Gaussian noise; see~\cite{Hoydis2013} for further details. Thus, $R_{jk}^{\rm{ul}}$ is a lower bound on the uplink ergodic capacity.

The second line of Eqn.~(\ref{sinr_ul}) shows that the uplink SINR takes the form of a generalized Rayleigh quotient. Therefore, a new multi-cell MMSE (M-MMSE) detector that maximizes the SINR in~(\ref{sinr_ul}) for given channel estimates is derived as:
\setcounter{equation}{15}
\begin{flalign} \label{detector}
& {\bf{g}}_{jk}^{\rm{M-MMSE}}\nonumber \\
& = \left( {\sum\limits_{l \in \cal L} {\sum\limits_{k = 1}^K \tau_{lk} \left({\hat {\bf h}}_{jlk}{\hat {\bf h}}_{jlk}^H +{\bf{C}}_{jlk} \right)} + {\sigma ^2}{{\bf{I}}_M}} \right)^{-1}{\hat {\bf h}}_{jjk}.
\end{flalign}
Furthermore, this detector minimizes the mean square error in estimating $x_{jk}$ as well \cite{tse2005fundamentals}:
\begin{equation}
\Eset \left\{ \left|{\hat x}_{jk} -x_{jk} \right|^2 \big|{\hat {\bf h}}_{\left(j \right)}\right\}.
\end{equation}
By plugging~(\ref{errormatrix2}) and~(\ref{estimation2}) into~(\ref{detector}), the M-MMSE detector can also be expressed as
\begin{equation} \label{detector2}
{\bf{g}}_{jk}^{\rm{M-MMSE}}= {\left( {{{{\hat {\bf H}}}_{\mathcal{V},j}}{{\bf{\Lambda }}_j}{\hat {\bf H}}_{\mathcal{V},j}^H + \left( {{\sigma ^2} + {\varphi_j}} \right){{\bf{I}}_M}} \right)^{ - 1}}{\hat {\bf h}}_{jjk},
\end{equation}
where ${{\bf{\Lambda }}_j} = \sum\limits_{l \in L} {\sum\limits_{k = 1}^K \tau_{lk} p_{lk}d_j^2\left({\bf z}_{lk}\right) } {{\bf{e}}_{{i_{lk}}}}{\bf{e}}_{{i_{lk}}}^H$. Apparently, ${\bf {\Lambda}}_j$ is a diagonal matrix and the $i$th diagonal element ${\lambda _{ji}}$ depends on the large scale fading, the pilot power and the payload power of users that use the $i$th pilot sequence in $\cal V$. The scalar ${\varphi_j}$ is
\begin{equation}
{\varphi_j} = \sum\limits_{l \in {\cal L}} {\sum\limits_{k = 1}^K {\tau_{lk}d_j\left({\bf z}_{lk}\right) \left(1-p_{lk}d_j\left({\bf z}_{lk}\right) \alpha_{ji_{lk}}B \right) }},
\end{equation}
where $\alpha_{ji_{lk}}$ is defined in~(\ref{temp}).

To elaborate the advantages of our M-MMSE scheme, we compare it with related work. First, the conventional single-cell MMSE (S-MMSE) detector from~\cite{Hoydis2013,Guo2014,Krishnan2014} is
\begin{equation} \label{smmse}
{\bf{g}}_{jk}^{\rm{S-MMSE}} = \left(  {\sum\limits_{m = 1}^K \tau_{jm}{{\hat {\bf h}}_{jjm}{\hat {\bf h}}_{jjm}^H} } + {\bf Z}_j + {\sigma ^2}{{\bf{I}}_M} \right)^{-1}{\hat {\bf h}}_{jjk},
\end{equation}
where inter-cell interference is either ignored by setting ${\bf Z}_j = \bf 0$ or only considered statistically as with
\begin{equation} \label{z}
{\bf Z}_j = \Eset\left\{ \sum\limits_{m=1}^K \tau_{jm}{{\tilde {\bf h}}_{jjm}{\tilde {\bf h}}_{jjm}^H} + \sum\limits_{l \ne j}\sum\limits_{m=1}^K \tau_{jm}{{ {\bf h}}_{jlm}{{\bf h}}_{jlm}^H}  \right\}.
\end{equation}

Notice that the S-MMSE detector only utilizes the $K$ estimated channel directions from within the serving cell, and treats directions from other cells as uncorrelated noise. Our M-MMSE detector, however, utilizes all the $B$ available estimated directions in ${\hat {\bf H}}_{{\cal V},j}$ so that BS $j$ can actively suppress also parts of inter-cell interference when $B > K$. Therefore, our detector can actually maximize the SINR in~(\ref{sinr_ul}), while S-MMSE can only do this in single-cell cases. The M-MMSE scheme can be seen as a coordinated beamforming scheme, but we stress that there is no signaling between the BSs since BS $j$ can estimate the channels ${\hat {\bf H}}_{{\cal V},j}$ directly from the uplink pilots. Moreover, the long-term channel statistics (like the channel attenuation and the pilot allocation) of all users to each BS can be obtained by using the uplink control channel, and hence does not necessarily require BS cooperation. Therefore, the M-MMSE scheme is fully scalable.

Compared with the multi-cell MMSE schemes proposed in~\cite{Guo2013} and~\cite{Hien2012}, our M-MMSE detector is more general and practical. To begin with, any pilot reuse policy and power control are supported in our scheme, which allows for an analysis based on a more flexible and practical network deployment. As shown in~\cite{bjornson2014massive}, non-universal pilot reuse is a reliable way to suppress pilot contamination and achieve high spectral efficiency in massive MIMO. In addition, as shown later on, it is with the larger pilot reuse factors that the M-MMSE achieves large gains over S-MMSE. Furthermore, the uplink detector in~\cite{Guo2013} is based on the unrealistic assumption that perfect CSI is known at the BS, while imperfect channel estimation is accounted for in our detector. Thus the performance gains provided by our detector are achievable in practical systems.

\section{Asymptotic Analysis} \label{asymptotic analysis}
In this section, performance analysis is conducted for the proposed M-MMSE detector. The SINR in~(\ref{sinr_ul}) is stochastic since it depends on the estimated channels in each block. Hence, the SE in~(\ref{rate_ul}) cannot be computed in closed form. We compute a deterministic equivalent of the SINR that is asymptotically tight. The large-system limit is considered, where $M$ and $K$ go to infinity while keeping ${K \mathord{\left/  {\vphantom {K M}} \right. \kern-\nulldelimiterspace} M} $ finite. In what follows, the notation $M \to \infty$ refers to $K$, $M \to \infty$ such that $\lim {\sup _M}{K \mathord{\left/  {\vphantom {K M}} \right. \kern-\nulldelimiterspace} M} < \infty $ and $\lim {\inf _M}{K \mathord{\left/  {\vphantom {K M}} \right. \kern-\nulldelimiterspace} M} >0$.\footnotemark{ }Since $B$ scales with $K$ for a fixed $\beta$, $\lim {\sup _M}{B \mathord{\left/  {\vphantom {K M}} \right. \kern-\nulldelimiterspace} M} < \infty $ and $\lim {\inf _M}{B \mathord{\left/  {\vphantom {K M}} \right. \kern-\nulldelimiterspace} M} >0$ also hold for $B$. The results should be understood in the way that, for each set of system dimension parameters $M$, $K$ and $B$, we provide a deterministic equivalent expression for the SINR, and the expression is a tight approximation as $M$, $K$ and $B$ grow large. The expression is ``deterministic'' because it only depends on the large-scale fading, pilot allocation and power control, while the SINR in~(\ref{sinr_ul}) depends also on the stochastic small-scale fading realizations.  In what follows, the notation $\xrightarrow[M \to \infty]{a.s.}$ denotes almost sure convergence of a stochastic sequence.
\footnotetext{The limit superior of a sequence $x_n$ is defined by $\lim {\sup _n}{x_n}\triangleq \mathop {\lim }\limits_{n \to \infty } \left( {\sup \left\{ {{x_m}:m \geqslant n} \right\}} \right)$; the limit inferior is defined as $\lim{\inf_n}{x_n}\triangleq \mathop {\lim }\limits_{n \to \infty } \left( {\inf \left\{ {{x_m}:m \geqslant n} \right\}} \right)$.}

Before we continue with our performance analysis, two useful results from random matrix theory are first recalled. All vectors and matrices should be understood as sequences of vectors and matrices of growing dimensions.

\subsection{Useful theorems} \label{theorems}
\begin{Theorem} (Theorem 1 in \cite{Wagner2012}): \label{theorem1}
Let ${\bf D} \in \Cset ^{M \times M}$ be deterministic and ${\bf H} \in \Cset ^{M \times B}$ be random with independent column vectors ${\bf h}_b \sim {\cal {CN}} \left(0, \frac{1}{M}{\bf R}_b \right)$. Assume that $\bf D$ and the matrices ${\bf R}_b \left( b=1,...,B\right)$, have uniformly bounded spectral norms (with respect to $M$). Then, for any $\rho > 0$,
\begin{equation}
\frac{1}{M} {\rm{tr}}\left({\bf D} \left( {\bf {HH}}^H +\rho {\bf I}_M \right)^{-1}\right) - \frac{1}{M} {\rm{tr}}\left({\bf D}{\bf T}\left( \rho \right)\right) \xrightarrow[M \to \infty]{a.s.} 0,
\end{equation}
where ${\bf T}\left( \rho \right) \in \Cset^{M \times M}$ is defined as
\begin{equation}
{\bf T}\left( \rho \right) = \left( \frac{1}{M}\sum\limits_{b=1}^B \frac{{\bf R}_b}{1+\delta_b \left( \rho\right)} +\rho{\bf I}_M\right)^{-1}
\end{equation}
and the elements of ${\bm {\delta }}\left( \rho  \right) \buildrel \Delta \over = {\left[ {{\delta _1}\left( \rho  \right),...,{\delta _B}\left( \rho  \right)} \right]^T}$ are defined as $\delta_b \left(\rho \right)=\lim_{t \to \infty}\delta_b^{\left(t \right)}\left( \rho \right), b=1,...,B$, where
\begin{equation}
\delta_b^{\left(t \right)}\left( \rho \right) = \frac{1}{M} {\rm{tr}} \left( {\bf R}_b \left( \frac{1}{M} \sum\limits_{j=1}^B \frac{{\bf R}_j}{1+ \delta_j^{\left(t-1\right)} \left(\rho \right)} +\rho{\bf I}_N\right)^{-1}\right)
\end{equation}
for $t=1,2,\ldots,$ with initial values $\delta_b^{\left(0 \right)} = 1/\rho$ for all $b$.
\end{Theorem}
\begin{Theorem}(see \cite{Wagner2012}) \label{theorem2}
Let ${\bf {\Theta}} \in \Cset^{M \times M}$ be Hermitian nonnegative definite with uniformly bounded spectral norm (with respect to $M$). Under the same conditions for $\bf D$ and $\bf H$ as in Theorem~\ref{theorem1},
\begin{gather}
\frac{1}{M} {\rm{tr}}\left({\bf D} \left( {\bf {HH}}^H + \rho {\bf I}_M \right)^{-1} {\bf \Theta}\left( {\bf {HH}}^H + \rho {\bf I}_M \right)^{-1} \right) \nonumber \\
- \frac{1}{M} {\rm{tr}}\left({\bf D}{\bf T}'\left( \rho \right) \right) \xrightarrow[M \to \infty]{a.s} 0
\end{gather}
where ${\bf T}'\left( \rho \right) \in \Cset^{M \times M}$ is defined as
\begin{equation}
{\bf T}'\left( \rho \right) ={\bf T}\left( \rho \right) {\bf \Theta} {\bf T}\left( \rho \right) +{\bf T}\left( \rho \right) \frac{1}{M} \sum\limits_{b=1}^B \frac{{\bf R}_b \delta'_b\left(\rho \right)}{\left(1+ \delta_b\left(\rho \right)\right)^2}{\bf T}\left( \rho \right).
\end{equation}
${\bf T}\left( \rho \right)$ and ${\bm \delta}\left( \rho  \right)$ are given by Theorem~\ref{theorem1}, and ${\bm{\delta }}'\left( \rho  \right)= \left[{\delta}'_1\left( \rho  \right),...,{\delta}'_B\left( \rho  \right) \right]^T$ is calculated as
\begin{equation}
{\bm \delta}' \left( \rho  \right)= \left({\bf I}_B - {\bf J}\left( \rho \right) \right)^{-1} {\bf v}\left( \rho \right)
\end{equation}
where ${\bf J}\left( \rho \right)$ and $ {\bf v}\left( \rho \right)$ are defined as
\begin{equation}
\left[ {\bf J}\left( \rho \right)\right]_{bl} = \mathLarge {\frac{ \frac{1}{M} {\rm{tr}} \left({\bf R}_b {\bf T}\left( \rho \right) {\bf R}_l {\bf T}\left( \rho \right)\right)} {M \left(1+\delta_l\left( \rho \right) \right)^2 }},  1 \le b,l \le B \\
\end{equation}
\begin{equation}
\left[ {\bf v}\left( \rho \right)\right]_{b} = \frac{1}{M}{\rm{tr}}\left({\bf R}_b{\bf T}\left( \rho \right) {\bf \Theta}{\bf T}\left( \rho \right)\right),  1 \le b \le B.
\end{equation}
\end{Theorem}

\setcounter{TempEqCnt}{\value{equation}}
\setcounter{equation}{30}
\vspace{-2ex}
\begin{figure*}[ht]
\begin{equation} \label{sinr_ul_determ}
{{\bar \eta }_{jk}^{\rm{ul}}} = \frac{\tau_{jk} p_{jk} d_j^2\left({\bf z}_{jk} \right){\delta _{jk}^2}}{{\delta_{jk}^2\sum\limits_{\left( {l,m} \right) \ne \left( {j,k} \right),{i_{_{lm}}} = {i_{jk}}} {\tau_{lm} p_{lm} d_j^2\left({\bf z}_{lm} \right)}  + \sum\limits_{{i_{_{lm}}} \ne {i_{jk}}} { \tau_{lm} d_j\left({\bf z}_{lm} \right)\frac{\mu_{jlmk}}{M}}  + \frac{{{\sigma ^2}}}{M}{\vartheta^{''}_{jk}}}}
\end{equation}
\hrule
\end{figure*}
\setcounter{equation}{\value{TempEqCnt}}

\subsection{Deterministic Equivalents of the SINR in~(\ref{sinr_ul})} \label{deterministic sinr} 
In what follows, we derive the deterministic equivalent ${\bar \eta}_{jk}^{\rm{ul}}$ of ${\eta}_{jk}^{\rm{ul}}$ for the M-MMSE detector such that
\begin{equation}
{\bar \eta}_{jk}^{\rm{ul}}-{\eta}_{jk}^{\rm{ul}} \xrightarrow[M \to \infty]{a.s.} 0.
\end{equation}
\begin{Theorem} \label{theorem3}
For the uplink MMSE detector in~(\ref{detector2}), we have $\eta_{jk}^{\rm{ul}} - {{\bar \eta }_{jk}}^{\rm{ul}} \xrightarrow[M \to \infty]{a.s.}0$, where $\bar{\eta}_{jk}^{\rm{ul}}$ is given in~(31) on top of the next page
with
\begin{flalign}
& \delta_{jk}=\frac{1}{M} {\rm{tr}} \left({\tilde {\bf \Phi}}_{{\cal V},ji_{jk}} {{\bf T}}_j\right) & \nonumber
\end{flalign}
\begin{flalign}
\mu_{jlmk}&=\frac{1}{M}{\rm{tr}} \left({{{{\mathbf{ T}}}_{jk}^{'}}}\right) \nonumber \\
 \begin{split}
&- p_{lm} d_j\left({\bf z}_{lm}\right)\gamma_{ji_{lm}}\vartheta_{jlmk}^{'}\vartheta_{jlm} \frac{2+\gamma_{ji_{lm}} \vartheta_{jlm} }{ \left(1+ \gamma_{ji_{lm}} \vartheta_{jlm}\right)^2 } \quad \quad \quad\nonumber
 \end{split}
\end{flalign}
\begin{flalign}
&\vartheta_{jlm}=\frac{1}{M}{\rm{tr}}\left( {\tilde {\bf \Phi}}_{{\cal V},ji_{lm}}{\bf T}_j\right)& \nonumber
\end{flalign}
\begin{flalign}
&\vartheta_{jlmk}^{'}=\frac{1}{M}{\rm{tr}}\left( {\tilde {\bf \Phi}}_{{\cal V},ji_{lm}}{\bf T}_{jk}^{'}\right)& \nonumber
\end{flalign}
\begin{flalign}
&\vartheta_{jk}^{''}=\frac{1}{M}{\rm{tr}}\left( {\tilde {\bf \Phi}}_{{\cal V},ji_{jk}}{\bf T}_{jk}^{''}\right) &\nonumber
\end{flalign}
where
\begin{enumerate}
\item ${\bf T}_j = {\bf T}_j\left(\alpha \right)$ and ${\bm{\delta }}\left( \alpha \right) \buildrel \Delta \over = {\left[ {{\delta _1},...,{\delta _B}} \right]^T}$ are given by Theorem~\ref{theorem1} for $\alpha = \frac{\sigma^2 + \varphi_j}{M}$ and ${\bf R}_b =\gamma_{jb}{ \tilde {\bf \Phi}}_{{\cal{V}},jb}$.  
\item ${\bf T}_{jk}^{'} = {\bf T}_{jk}^{'}\left(\alpha \right)$ and ${\bm{\delta }}'\left( \alpha  \right)= \left[{\delta}'_1,...,{\delta}'_B \right]^T$ are given by Theorem~\ref{theorem2} for $\alpha = \frac{\sigma^2 +\varphi_j}{M}$, ${\bf \Theta}={\tilde{\bf \Phi}}_{{\cal V},ji_{jk}}$, and ${\bf R}_b =\gamma_{jb}{\tilde {\bf \Phi}}_{{\cal{V}},jb}$. 
\item ${\bf T}_{jk}^{''}={\bf T}_{jk}^{''} \left(\alpha \right)$ and ${\bm{\delta }}'\left( \alpha  \right)= \left[{\delta}'_1,...,{\delta}'_B \right]^T$ are given by Theorem~\ref{theorem2} for $\alpha = \frac{\sigma^2 + \varphi_j}{M}$, ${\bf \Theta}={\bf I}_M $, and ${\bf R}_b =\gamma_{jb}{\tilde {\bf \Phi}}_{{\cal{V}},jb}$.  
\end{enumerate}
\end{Theorem}
\noindent\emph{Proof:} The main idea behind the proof is to derive the deterministic equivalent of each term of the first line of~(\ref{sinr_ul}). Then the deterministic equivalent of~(\ref{sinr_ul}) is obtained as~(31). The full proof can be found in the Appendix B of~\cite{xueru2015}. \hfill{$\blacksquare$}

With the ${\bar \eta_{jk}}^{\rm{ul}}$ above, the ergodic SE in~(\ref{rate_ul}), after dropping the prelog factor, converges to ${\bar R}_{jk}^{\rm{ul}}=\log_2 (1+{\bar \eta_{jk}}^{\rm{ul}})$. Since ${\bar \eta_{jk}}^{\rm{ul}}$ does not depend on the instantaneous small-scale channel fading, $(1-\frac{B}{S}){\bar R}_{jk}^{\rm{ul}}$ is a large-scale approximation of the ergodic SE. As shown later, this approximation is even very accurate at small system dimensions. Furthermore, the approximation is easy to compute and allows for simple performance analysis.

\section{Simulation Results} \label{simulation}
In this section, we illustrate the accuracy and usefulness of the analytical contributions for a symmetric hexagonal network topology. We apply the classic 19-cell-wrap-around structure to avoid edge effects and guarantee the same performance for all cells. Each hexagonal cell has a radius of $r = 500$ meters, and is surrounded by 6 interfering cells in the first tier, and 12 in the second tier. To achieve a symmetric pilot allocation network, the pilot reuse factor can be $\beta \in \left\{1,3,4,7,...\right\}$.

User locations are generated independently and randomly in the cells by following uniform distributions, but the distance between each user and its serving BS is at least $0.14r$. For each user location ${\bf z} \in \Rset^{2}$, a classic pathloss model is considered, where the variance of channel attenuation is $d_j\left( {\bf z} \right) = \frac{C}{\left\| {\bf z} - {\bf b}_j\right\|^{\kappa}}$. Here ${\bf b}_j \in \Rset^{2}$ is the location of the BS in cell $j$, $\kappa$ is the pathloss exponent, and $\left\|\cdot \right\|$ denotes the Euclidean norm. $C>0$ is independent shadow fading with $10\log_{10}\left(C \right)\sim {\cal N}(0,\sigma^2_{sf})$. We assume $\kappa = 3.7$, $\sigma_{sf}^2 = 5$ and coherence block length $S= 300$.\footnotemark{} Orthogonal pilots are chosen from a $B \times B$ discrete Fourier transform unitary matrix.
\footnotetext{This coherence block can, for example, have the dimensions of $T_c =3\,{\rm{ms}}$ and $W_c = 100\,{\rm{kHz}}$.}

Statistical channel inversion power control is applied to both pilot and payload data transmission~\cite{bjornson2014massive}, i.e., $p_{lk} = \tau_{lk} =\frac{\rho}{d_l\left({\bf z}_{lk}\right)}$. Thus the average effective channel gain between users and their serving BSs is constant: $\Eset\{p_{lk}\left\|{\bf h}_{llk}\right\|^2\} = M\rho$, and the statistical per antenna received SNR of each user at its serving BS is $\rho/{\sigma^2}$. This is a simple but effective policy to avoid near-far blockage and, to some extent, guarantee a uniform user experience. In the simulation, $\rho/{\sigma^2}$ is set to 0 dB to allow for decent channel estimation accuracy.

\begin{figure}[h]
\centering
\scalebox{0.45}{\includegraphics*{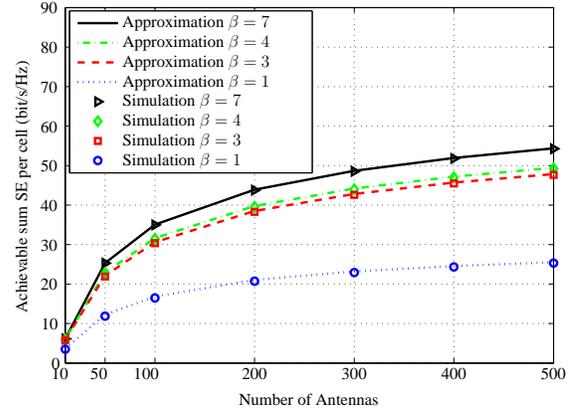}} 
\caption{Achievable sum SE as a function of the number of antennas $M$, for $\beta\in\left\{1,3,4,7 \right\}$ and $K=10$.}
\label{inst_determ}
\end{figure}

In the simulation, 10000 independent Monte-Carlo channel realizations for small scale fading are generated to numerically calculate the achievable SE in~(\ref{rate_ul}). The numerical result and its large-scale approximation from Theorem~\ref{theorem3} are shown in Fig.~\ref{inst_determ} for $K=10$ and different $M$. Fig.~\ref{inst_determ} shows that the achievable sum SE increases monotonically as $\beta$ grows, at least for $\beta \leq 7$. This is a result of the following two properties. Firstly, a larger $\beta$ results in a lower level of pilot contamination, contributes to a higher channel estimation accuracy, and thereby increases the system SE. Secondly, a larger $\beta$ also indicates more available estimated channel directions in the M-MMSE detector, thus a higher inter-cell interference suppression can be achieved. Fig.~\ref{inst_determ} also shows that the Monte-Carlo simulations and the large-scale approximations match well, even for relatively small $M$ and $K$.
\psfrag{Approximation beta=7 blablabla}{\Large {Approximation $\beta = 7$}}
\psfrag{data2}{\Large {Approximation $\beta = 4$}}
\psfrag{data3}{\Large {Approximation $\beta = 3$}}
\psfrag{data4}{\Large {Approximation $\beta = 1$}}
\psfrag{data5}{\Large {Simulation $\beta = 7$}}
\psfrag{data6}{\Large {Simulation $\beta = 4$}}
\psfrag{data7}{\Large {Simulation $\beta = 3$}}
\psfrag{data8}{\Large {Simulation $\beta = 1$}}
\psfrag{Number of Antennas}[][cb]{\Large {Number of Antennas}}
\psfrag{Achievable sum rate per cell (bit/s/Hz)}[][]{\Large{Achievable sum SE per cell (bit/s/Hz)}}
\psfrag{U=300}{}

\psfrag{10}[][]{\Large {10}}
\psfrag{50}[][]{\Large {50}}
\psfrag{100}[][]{\Large {100}}
\psfrag{200}[][]{\Large {200}}
\psfrag{300}[][]{\Large {300}}
\psfrag{400}[][]{\Large {400}}
\psfrag{500}[][]{\Large {500}}

\begin{figure}[H]
\centering
\scalebox{0.45}{\includegraphics*{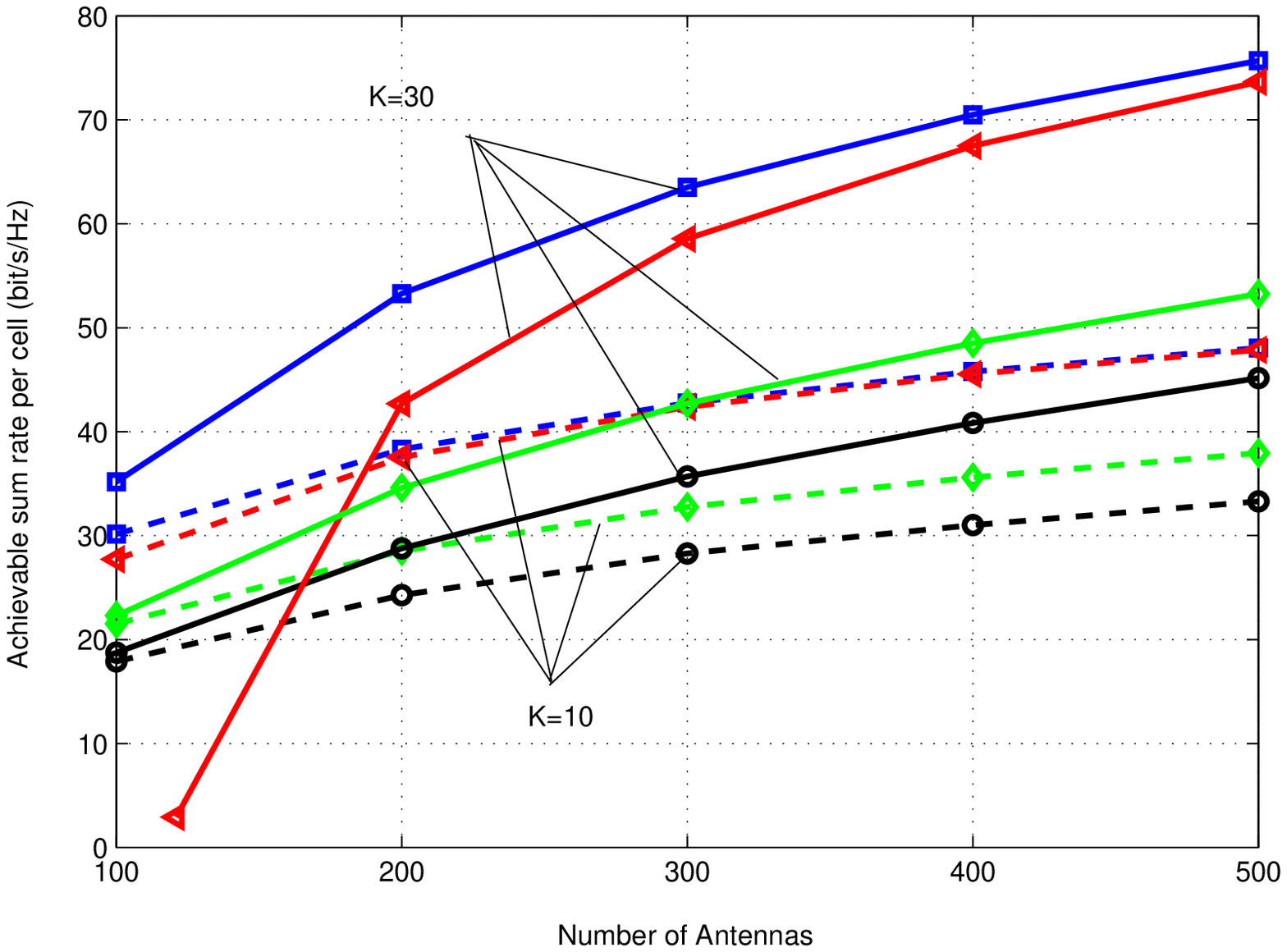}}
\caption{Achievable sum SE of M-MMSE (squares), M-ZF (triangles), S-MMSE (diamonds) and M-MF (circles) with $\beta =4$, $K=10$ and $K=30$.}
\label{sumrate_beta4}
\end{figure}

\begin{figure}[t]
\centering
\scalebox{0.45}{\includegraphics*{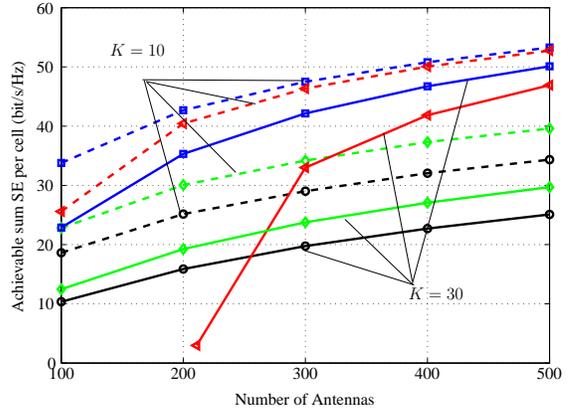}}
\caption{Achievable sum SE of M-MMSE (squares), M-ZF (triangles), S-MMSE (diamonds) and M-MF (circles) with $\beta =7$, $K=10$ and $K=30$.}
\label{sumrate_beta7}
\end{figure}

\psfrag{0}[][l]{\Large {0}}
\psfrag{10}[][l]{\Large {10}}
\psfrag{20}[][l]{\Large {20}}
\psfrag{30}[][l]{\Large {30}}
\psfrag{40}[][l]{\Large {40}}
\psfrag{50}[][l]{\Large {50}}
\psfrag{60}[][l]{\Large {60}}
\psfrag{70}[][l]{\Large {70}}
\psfrag{80}[][l]{\Large {80}}
\psfrag{90}[][l]{\Large {90}}
\psfrag{100}[][l]{\Large {100}}
\psfrag{120}[][l]{\Large {120}}
\psfrag{K=30}{\Large {$K=30$}}
\psfrag{K=10}{\Large {$K=10$}}
\vspace{-3ex}
\vspace{-1ex}
To show explicitly the advantages of our M-MMSE detection scheme, simulation results for the matched filter (MF) from~\cite{Marzetta10}, the multi-cell ZF (M-ZF) detector from~\cite{bjornson2014massive}, and the S-MMSE detector from~(\ref{smmse}) are provided for comparison. Notice that $M-\beta K>0$ is needed for the M-ZF scheme, thus the minimum value of $M$ for the M-ZF is $\beta K + 1$. Since Fig.~\ref{inst_determ} shows that $\beta=4$ and $\beta=7$ give the highest performance, we provide simulation results for $\beta=4$ and $\beta=7$ in Fig.~\ref{sumrate_beta4} and Fig.~\ref{sumrate_beta7}, respectively. The MF scheme always achieves the lowest performance since it does not actively suppress any inter-user interference. Compared with the S-MMSE, our M-MMSE achieves a notable SE gain and the advantage becomes more significant as $\beta$ and/or $K$ increases. For $\beta=4$ and $M = 200$, the SE of M-MMSE are 28\% and 56\% higher than those of S-MMSE for $K = 10$ and $K = 30$, respectively. When $\beta=7$, the gains increase to 40\% and 84\%, respectively. Notice that when $\beta=7$, $K=30$ brings lower achievable rates compared with $K=10$, due to the loss from a large pilot overhead. The advantage of the M-MMSE over the M-ZF is relatively small for $K=10$, but it becomes significant as $\beta$ and $K$ grow. Moreover, the M-ZF can sometimes achieve very low SE for small $M$, while our M-MMSE can always achieve good performance, with the same complexity as for the M-ZF. From the analysis above, it can be concluded that the proposed M-MMSE brings a very promising gain over single-cell processing and the M-ZF, and the gain becomes increasingly significant as $\beta$ and/or $K$ grow.

\psfrag{K=30}{\Large {$K=30$}}
\psfrag{K=10}{\Large {$K=10$}}
\psfrag{Number of Antennas}[][cb]{\Large {Number of Antennas}}
\psfrag{Achievable Rate (bit/s/Hz)}[][]{\Large{Achievable Rate (bit/s/Hz)}}
\psfrag{500}[][]{\Large {500}}
\psfrag{200}[][]{\Large {200}}
\psfrag{300}[][]{\Large {300}}
\psfrag{400}[][]{\Large {400}}
\psfrag{100}[][]{\Large {100}}
\psfrag{120}[][]{\Large {120}}
\psfrag{140}[][]{\Large {140}}
\psfrag{160}[][]{\Large {160}}
\psfrag{data27}{\Large {Multicell MF}}
\psfrag{data37}{\Large {Multicell ZF}}
\psfrag{data47}{\Large {Singlecell MMSE}}
\psfrag{2}[][l]{\Large {2}}
\psfrag{3}[][l]{\Large {3}}
\psfrag{4}[][l]{\Large {4}}
\psfrag{5}[][l]{\Large {5}}
\psfrag{6}[][l]{\Large {6}}
\psfrag{7}[][l]{\Large {7}}
\vspace{-2ex}

\vspace{-2ex}
\section{Conclusions} \label{conclusions}
In this paper, a multi-cell MMSE detector is proposed and a tight deterministic equivalent SINR expression is derived in the large-system limit. Compared with the conventional single-cell MMSE scheme, that only utilizes the estimated directions from within the serving cell, the proposed multi-cell MMSE scheme utilizes all channel directions that can be estimated locally at the BS to suppress the inter-cell interference. Numerical results show that the proposed multi-cell MMSE brings very promising sum SE gains over the single-cell MMSE and the multi-cell ZF. Since imperfect estimated CSI is accounted for in our scheme, the gains obtained are likely to be achievable in practical systems. The M-MMSE scheme is the new state-of-the-art method for massive MIMO detection and is hard to beat since it maximizes the SINR under very general conditions. Furthermore, the deterministic equivalent is shown to be accurate even for relatively small system dimensions.
\vspace{-2ex}
\bibliographystyle{IEEEtran}
\linespread{1.0}\selectfont
\bibliography{mmse_gc_ul}

\end{document}